\documentclass[preprint]{aastex}
\usepackage{graphicx}
\usepackage{subfigure}
\usepackage{natbib}
\usepackage{amsmath}
\usepackage[usenames,dvipsnames]{color}

\newcommand{\mycitet}[1]{\citet{#1}}
\newcommand{\mycitep}[1]{\citep{#1}}
\newcommand{\mycitealt}[1]{\citealt{#1}}

\newcommand{\myfig}[1]{Fig. \ref{#1}}
\newcommand{\intd}[1]{\ensuremath{\,\mathrm{d}#1}}

\title{A dual-mask coronagraph for observing faint companions to binary stars}
\author{Eric Cady$^{1}$, Michael McElwain$^{2,3,4,5}$, N. Jeremy Kasdin$^{2}$, Christian Thalmann$^{6}$}

\begin{abstract}
Observations of binary stars for faint companions with conventional coronagraphic methods are challenging, as both targets will be bright enough to obscure any nearby faint companions if their scattered light is not suppressed.  We propose coronagraphic examination of binary stars using an apodized pupil Lyot coronagraph and a pair of actively-controlled image plane masks to suppress both stars simultaneously.  The performance is compared to imaging with a band-limited mask, a dual-mask Lyot coronagraph and with no coronagraph at all.  An imaging procedure and control system for the masks are also described.
\end{abstract}

\keywords{binaries: general, instrumentation: high angular resolution, planetary systems, techniques: high angular resolution}

\begin{document}
\maketitle
\footnotetext[1]{Jet Propulsion Laboratory, California Institute of Technology, Pasadena, CA 91109, USA.  Contact author: eric.j.cady@jpl.nasa.gov}
\footnotetext[2]{Dept. of Mechanical and Aerospace Engineering, Princeton University, Princeton, NJ 08544, USA.}
\footnotetext[3]{Princeton University Observatory, Princeton, NJ 08544, USA}
\footnotetext[4]{Department of Astrophysical Sciences, Princeton University, Princeton, NJ 08544, USA}
\footnotetext[5]{NASA Goddard Space Flight Center, Exoplanets and Stellar Astrophysics Laboratory, Greenbelt, MD 21230, USA}
\footnotetext[6]{Max Planck Institute for Astronomy, K\"onigstuhl 17, D-69117 Heidelberg, Germany}
\pagebreak

\section{Introduction}

Although a large fraction of nearby stars occur in multiple systems, direct detection exoplanet searches have generally selected against binary targets because a secondary star complicates observations and limits the detection sensitivity.  In addition to observational complications, binary systems were traditionally assumed to be a hostile environment for the formation and evolution of planetary systems.  However, direct imaging of stars known to harbor planets detected by radial velocities has revealed several binary star systems \mycitep{Pat02, Egg09}. Protoplanetary disks, which form the material basis for planet formation, are observed in both circumstellar and circumbinary configurations in binary systems \mycitep{Rod96, Tri07}, and the growth and settling of dust grains is common in binary star systems \mycitep{Pas08}.  Theoretical numerical simulations successfully model the evolution of protoplanets \mycitep{Pie07}, terrestrial planets \mycitep{Qui06}, giant planets \mycitep{Pie08} and brown dwarfs \mycitep{Jia04} in circumbinary disks.  Furthermore, theoretical models also permit planet formation through gravitational instability in binary systems.  Planets formed in this way have large separations from their stars and are therefore particularly important targets for direct imaging.  The influence of a secondary star can in some cases prevent a collapse through tidal heating \mycitep{May05} or trigger the collapse for an otherwise stable disk \mycitep{Bos06}, leaving a characteristic imprint on the demographics of planets in binary systems, the observation of which would provide an invaluable test for planet formation theory.  A secondary star can also act as a source of angular momentum for a circumbinary planet and either scatter or tidally push the planet into wider orbits, thus enriching the expected population of wide-orbit stars \mycitep{Nel03, Ver04, Hol99, Kle00}.

Direct detection techniques are critical for investigating extrasolar planets on large ($>$ 10 AU) orbital separations, and the first direct detections of extrasolar planets have confirmed that planets do exist in these orbits (e.g., \mycitealt{Mar08,Kal08,Lag09}).  High contrast instrumentation and direct detection techniques are being developed to probe planet formation and the evolution of planetary systems.  In order to achieve high contrast ($>$ 10$^{-6}$) at small angular separations ($<$ 1$\arcsec$), one must control the diffracted light from the host star in the image plane.  However, the majority of coronagraphs are designed to remove light from single host stars.  If the secondary star lies close to the target star and it is not suppressed, the secondary star's light will overwhelm the signal from any faint companions.  The peak of the G dwarf companion distribution is near 30 AU \mycitep{Duq91}, and while also accounting for random orbital inclinations and phases on the sky, the vast majority of nearby binary stars ($<$ 100 pc) targeted by high contrast imaging surveys would benefit from having a specialized coronagraph.

Coronagraphs capable of simultaneously blocking the light from binary stars can be grouped into two categories: coronagraphs with linear masks, and coronagraphs with dual circular masks.  Each type has advantages and disadvantages; in the following, we present examples of each and discuss their performance in the presence of a typical low-order wavefront noise following an adaptive optics system.  In particular, we present a dual-mask design based on an APLC which allows for a small inner working angle and a large discovery space even for an obstructed aperture, and which minimizes cross-talk between the masks.  We also address manufacturing and implementation concerns such as obstructed apertures, field rotation, manufacturing, and mask placement, and outline a control scheme for maintaining alignment of the system even in the presence of atmospheric errors.

\section{Types of Coronagraphs for High Contrast Imaging}

The general class of Lyot-type coronagraphs have a common structure to the optical design.  The primary components of a Lyot coronagraph consist of an initial pupil equivalent to the pupil of the telescope, potentially with an apodization; a focal plane mask, which can be hard-edged or vary in amplitude and phase; and a Lyot stop at the reimaged pupil, which is hard-edged and often (but not always) undersized with respect to the telescope pupil.  Lyot's original design \mycitep{Lyo39} was equipped with hard-edged masks in all planes; however, subsequent variants include apodized pupil Lyot coronagraphs (APLCs) \mycitep{Sou03}, band-limited coronagraphs (BLCs) \mycitep{Kuc02}, four-quadrant phase mask coronagraphs (4QPMs) \mycitep{Rou00}, and vortex coronagraphs (VCs) \mycitep{Foo05, Maw10}.  \mycitet{Guy06} provides a detailed analysis of comparative coronagraphy.

For the specialized case of binary stars, these coronagraphs split into two general categories: (1) coronagraphs with a single focal plane mask which is oriented to block both stars simultaneously, and (2) coronagraphs with two masks, one placed to block each star.

\subsection{Single-mask case}

The main purpose behind using a single mask is simplicity---it avoids additional mechanisms and complexity by having a single fixed mask.  In exchange, this mask will tend to block out more of the field of the view than just the target stars, obscuring edge-on systems in particular.

These masks work by using a separable apodization $A(x, y) = A(x)A(y)$, generally with $A(y) = 1$.  The mask is then oriented so that $y$-axis is aligned with the two stars, and both will be suppressed identically.  Coronagraphs that could be considered for this approach include the BLC---in fact, this has been demonstrated on-sky \mycitep{Cre10}---and an APLC designed for a linear mask, following \mycitet{Aim02}.

\subsection{Dual-mask case}

The rationale behind using two masks in the image plane is to block only the regions of the field of view containing the starlight, so the remainder of the field can be investigated as discovery space.  However, the discovery space of the coronagraphic mask may not be symmetric (as is the case for the BLC), which could lead to significant preference for certain orbital inclinations.  Observations of T-Tauri stars, for example, indicate circumstellar disks in binary systems tend to be parallel to each other \mycitep{Mon07}. Assuming that all spatial orientations are randomly distributed on the sky, the abundance of orbits with a specific inclination angle is proportional to the cosine of the inclination.   This represents a significant bias towards encountering edge-on binary systems and suggests that the binary orientation direction is a favorable place to find a planetary companion, assuming the planet formed in the disk and was not scattered far from the invariable plane.  A dual-mask coronagraph enables sensitivity along this orientation, while the BLC completely obscures planets along that axis.  Therefore, the BLC has a strong selection preference for observing face-on systems, where this effect is minimized, but the dual-mask coronagraph has no such requirement.  On the other hand, a dual-mask coronagraph system is significantly more complex, as the two masks must be moved in the image plane to coincide with the target stars.

\begin{figure}
\begin{center}
\includegraphics[width=6.5in]{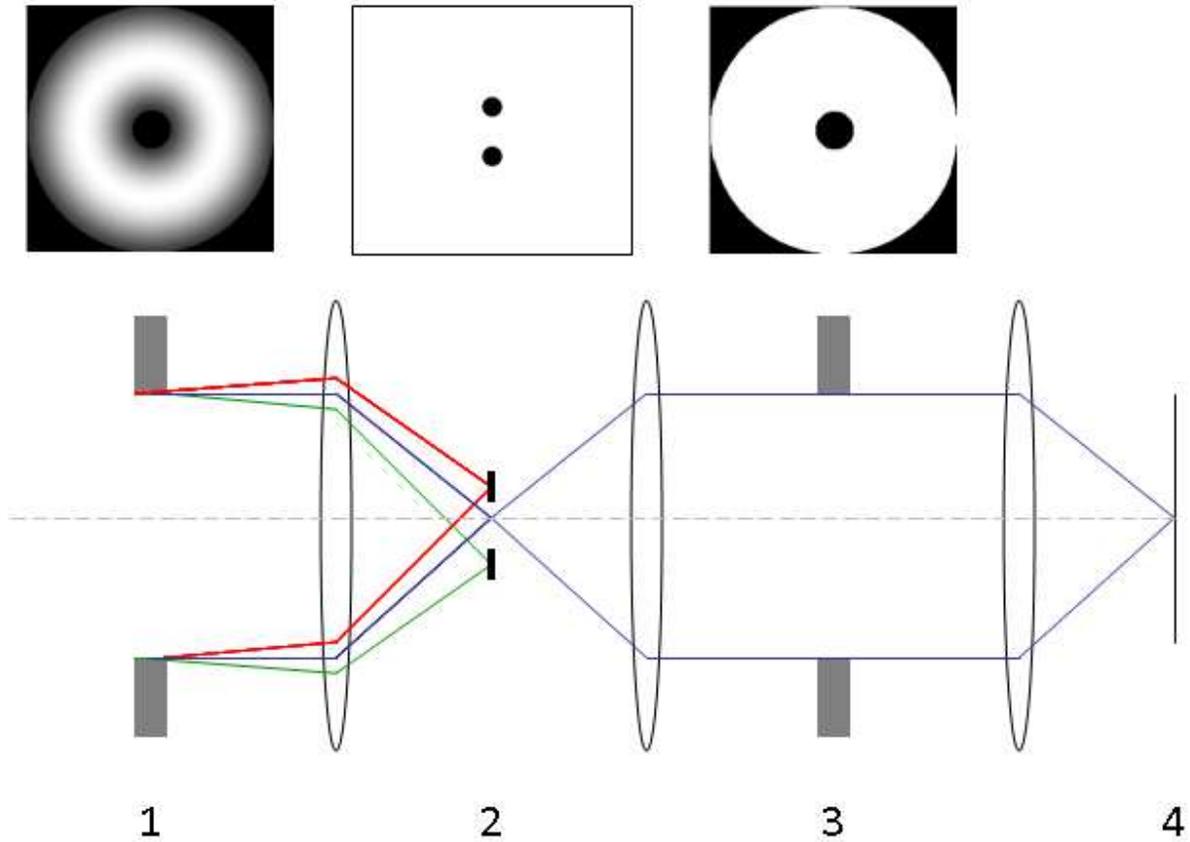}
\end{center}
\caption{APLC optical components for binary observations.  Plane $1$ has an apodizer and the pupil, shown face-on at top; plane $2$ has the dual image plane masks, shown face-on above the plane; plane $3$ is the pupil again, shown face-on above; plane $4$ represents the science camera image plane.  The red and green rays show the path of the starlight from the binary pair through the system, while the blue ray shows the path of light from the planet.} \label{fig:bscdiagram}
\end{figure}

A source located at an angular separation of $\psi_1$ with respect to the optical axis produces a tilted wavefront at the aperture of the telescope; without loss of generality, we can place this source along the $x_1$-axis with unit intensity to give the time-averaged electric field $\Psi_{1}(x_1,y_1; \psi_1)$ following the first pupil plane:
\begin{equation} \label{tilt1}
    \Psi_{1}(x_1,y_1; \psi_1) = P(x_1,y_1) e^{[-2 \pi i  x_1 \sin{\psi_1}/ \lambda]}
\end{equation}
with $P(x_1,y_1)$ the pupil function of the telescope and $\lambda$ the wavelength under consideration. The electric field at the subsequent image plane ($\Psi_{2}(x_2,y_2; \psi_1)$) can be written:
\begin{align}
    \Psi_{2}(x_2,y_2; \psi_1) &= M(x_2, y_2) \times \frac{1}{i \lambda f} \int_{-\infty}^{\infty} \int_{-\infty}^{\infty} e^{-\frac{2 \pi i}{\lambda f} (x_1 x_2 + y_1 y_2)} P(x_1,y_1) e^{-\frac{2 \pi i  x_1 \sin{\psi_1}}{\lambda}} \intd{x_1} \intd{y_1} \\
    &= M(x_2, y_2) \times \frac{1}{i \lambda f} \int_{-\infty}^{\infty} \int_{-\infty}^{\infty}  P(x_1,y_1) e^{-\frac{2 \pi i}{\lambda f} (x_1 [x_2 + f \sin{\psi_1}] + y_1 y_2)} \intd{x_1} \intd{y_1}  \label{tilt2}
\end{align}
with $f$ the focal length of the telescope and $M(x_2, y_2)$ the focal plane mask.  In the on-axis case,
\begin{align}
    \Psi_{2}(x_2,y_2; 0) &= M(x_2, y_2) \times \frac{1}{i \lambda f} \int_{-\infty}^{\infty} \int_{-\infty}^{\infty}  P(x_1,y_1) e^{-\frac{2 \pi i}{\lambda f} (x_1 x_2 + y_1 y_2)} \intd{x_1} \intd{y_1},  \label{tilt2a}
\end{align}
and so the shift by $f \sin{\psi_1}$ from the tilt of the incident wavefront creates a misalignment with the mask.  We can compensate by moving the mask by $-f \sin{\psi_1}$:
\begin{align}
    \Psi_{2s}(x_2,y_2; \psi_1) &= M(x_2 + f \sin{\psi_1}, y_2) \times \frac{1}{i \lambda f} \int_{-\infty}^{\infty} \int_{-\infty}^{\infty}  P(x_1,y_1) e^{-\frac{2 \pi i}{\lambda f} (x_1 [x_2 + f \sin{\psi_1}] + y_1 y_2)} \intd{x_1} \intd{y_1} \label{tilt3a} \\
    &= \Psi_{2}(x_2 + f \sin{\psi_1},y_2; 0) \label{tilt3b}
\end{align}
so the system with an off-axis source provides identical suppression at $x_2 = -f \sin{\psi_1}$ as does an on-axis mask and star.  (The image plane masks behave achromatically in this configuration, where the wavelength does not change the magnitude of the shift of the PSF in the image plane.)  If the two stars are located at $\pm \psi_1$ from the optical axis, then, the two masks would be placed at $\pm f \sin{\psi_1}$, and would suppress both stars simultaneously.  A schematic of this system is shown in \myfig{fig:bscdiagram}, for a simple telescope with circular symmetry.

Coronagraphs that could in theory be considered for this dual-mask approach include nearly every Lyot-type coronagraph with a radially-symmetric focal-plane mask, including Lyot coronagraphs, APLCs, BLCs with radially-symmetric masks \mycitep{Kuc03}, 4QPMs, and VCs.  Unfortunately, this simple analysis ignores one large difficulty: not only does mask A block the core of star A, but mask B will suppress a portion of the sidelobes of star A, which then will scatter light across the image plane, limiting the contrast that can be achieved.  Thus, only a coronagraph with the lowest sidelobes at the first image plane is likely to prove promising for a binary system.  We investigate the APLC because it includes an apodization at the entrance pupil that reshapes the PSF.  As a result, the APLC has optimally low sidelobes \mycitep{Sou03} and will minimize the interaction between the two masks.

\subsection{Additional design considerations}

Most existing telescopes have obstructed apertures, and the effects of secondaries and their support structures on the coronagraphs must be considered in the modeling of the optical performance.  In many cases, the effects of these optical obstructions can be mitigated by carefully designing the coronagraph.  For example, Lyot stops can be designed to block the diffraction from aperture obstructions \mycitep{Siv05}, spider removal plates can remove the spiders from the image \mycitep{Loz09}, and for APLCs, the apodization can be optimized for arbitrary apertures \mycitep{Sou09}.  However, the reduced performance should be taken into account when designing these coronagraphs.  \mycitet{Cre10} makes binary star observations with the unobstructed subaperture \mycitep{Ser07} on the Palomar Observatory's 200-inch Hale Telescope, neatly avoiding these coronagraphic complications, but only utilizing $\sim9\%$ of the collecting area of the telescope.

A further difficulty arises from the field rotation, which causes the binary system to rotate with respect to the telescope pupil for telescopes in altitude-azimuth mounts.  In high contrast observations, this characteristic can be employed for Angular Differential Imaging (ADI) \mycitep{Mar06}.  The major consequence of this is that the coronagraph must be able to maintain its performance regardless of the orientation of the aperture with respect to the image plane mask, including any obstructions.  For an unobstructed circular aperture, the circular symmetry of the pupil renders this trivial.  The performance of the system is also dependent on the aberrations introduced by the atmosphere and the ability of the adaptive optics (AO) system to correct these wavefront errors.

\section{Simulated performance} 

\begin{figure}
\begin{center}
\includegraphics[width=3.1in]{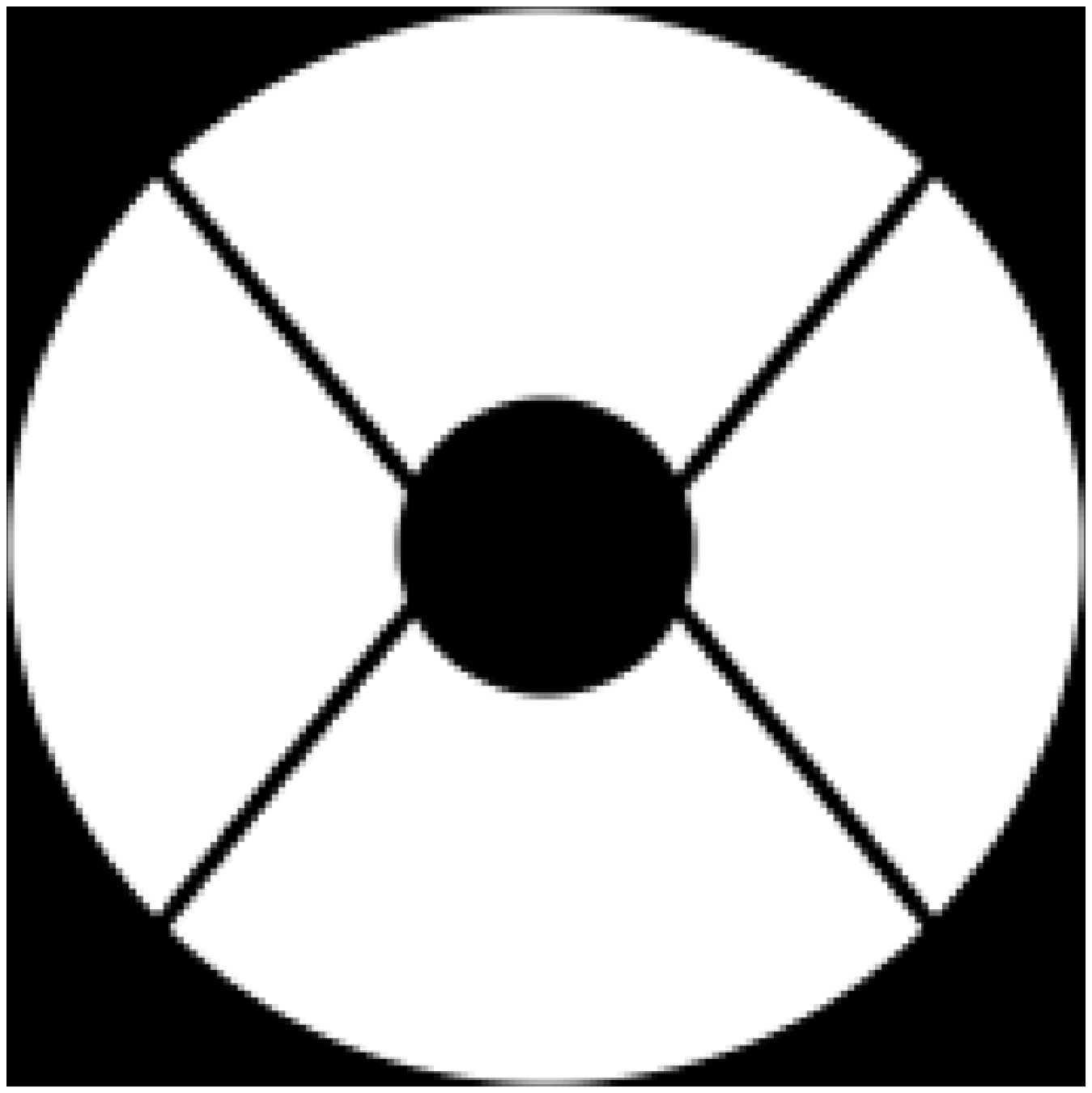}
\caption{The pupil of Subaru telescope.}
\label{subaruPupil}
\end{center}
\end{figure}

To examine the performance of the coronagraphic options discussed above, we simulate their performance on a realistic system.  Here, we use parameters of the Subaru Telescope and HiCIAO \mycitep{Hod08} for simulation purposes: the pupil is a $8.2$m diameter circular aperture with an IR secondary $1.265$m diameter circular obscuration concentric with the primary mirror \mycitep{Usu03}.  The secondary is held in place with four spiders, as shown in \myfig{subaruPupil}.  From a numerical perspective, we make extensive use of the Fast Lyot algorithm described in \mycitet{Sou07} for efficient propagation.

The simulated field contains a binary system and a faint companion.  The system is simulated at $1.65\mu$m for a 10s exposure, using simulated residual wavefronts from the atmosphere following the AO188 adaptive optics system \mycitep{Min10} at 0.5ms intervals, with 200nm of residual static rms wavefront error.  The two stars are placed at $\pm 500$\,mas along the $y$-axis, with intensity normalized to 1; the companion is placed at $-1000$\,mas along the $x$-axis and is attenuated to an intensity of $10^{-4}$ for clarity.  We choose a 1'' separation of the binary companions because closer pairs interfere with the wavefront sensing capability of the AO188 system (S. Egner, private communication) and thus are unlikely to have a well-corrected field.

\begin{figure}
\begin{center}
\subfigure[The binary system with no coronagraph.]{\label{m1} \includegraphics[width=3.1in]{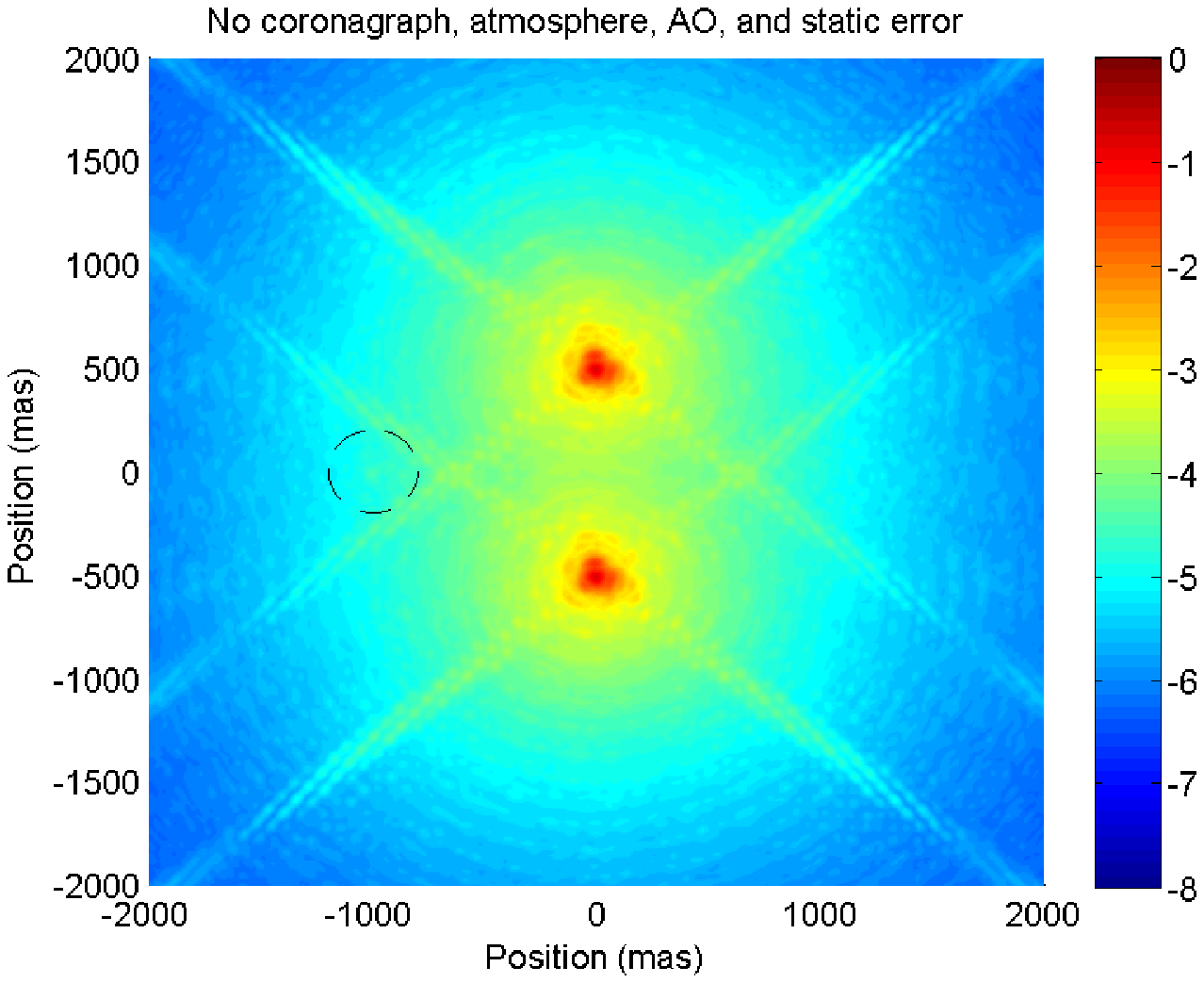}}
\subfigure[The binary system with a Lyot coronagraph.]{\label{m2} \includegraphics[width=3.1in]{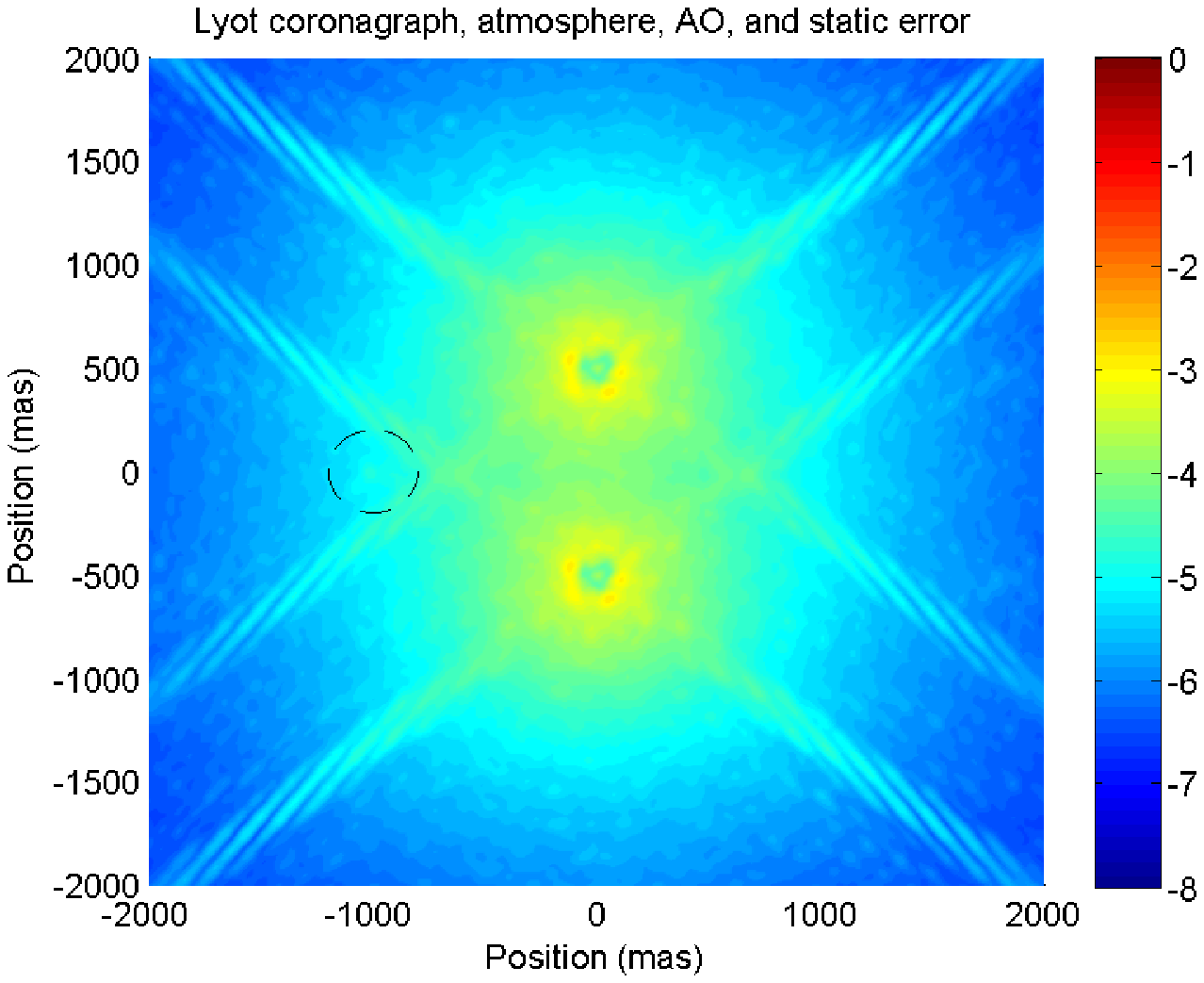}}
\subfigure[The binary system with an apodized pupil Lyot coronagraph.]{\label{m3b} \includegraphics[width=3.1in]{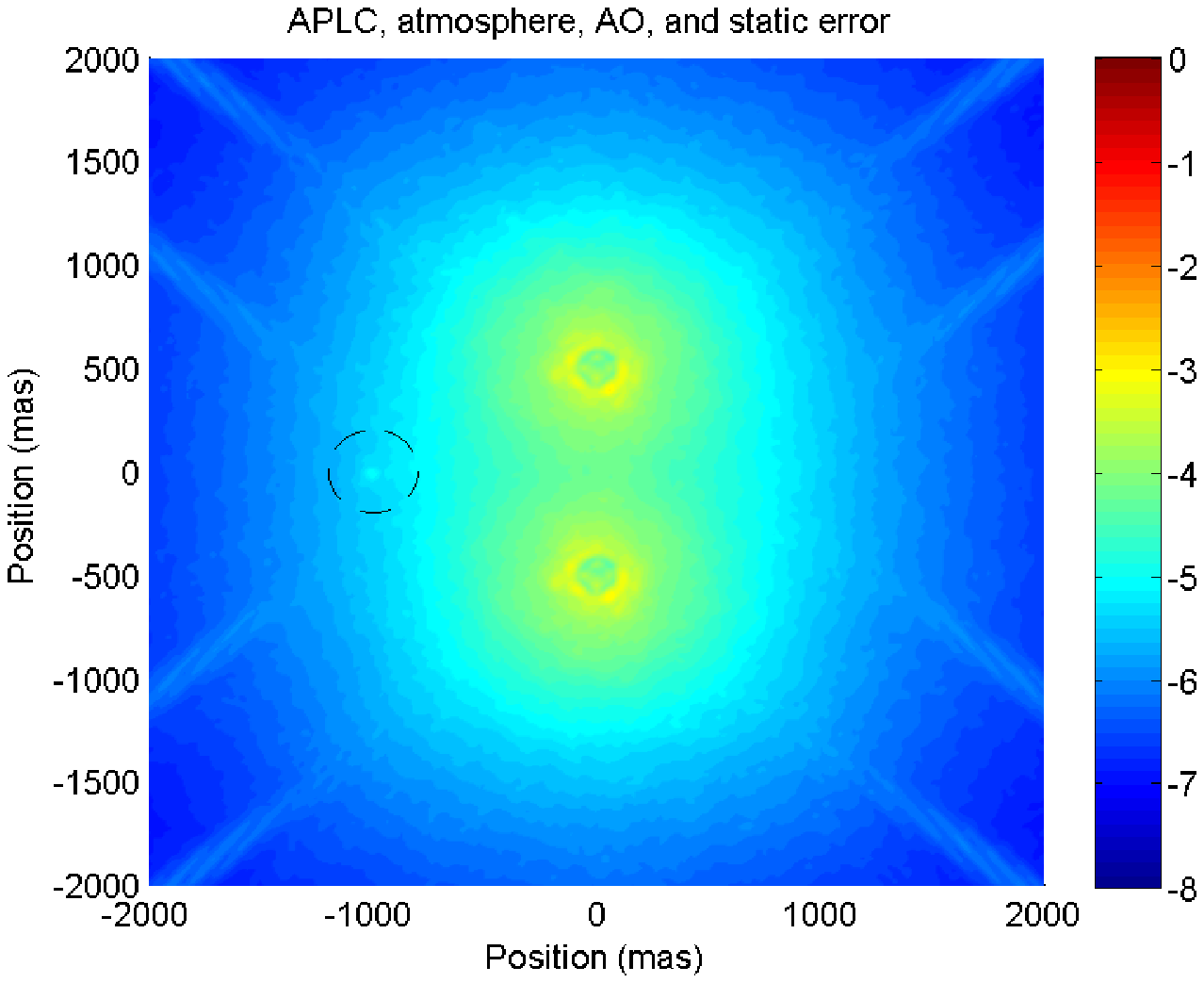}}
\subfigure[The binary system with a band-limited coronagraph.]{\label{m4b} \includegraphics[width=3.1in]{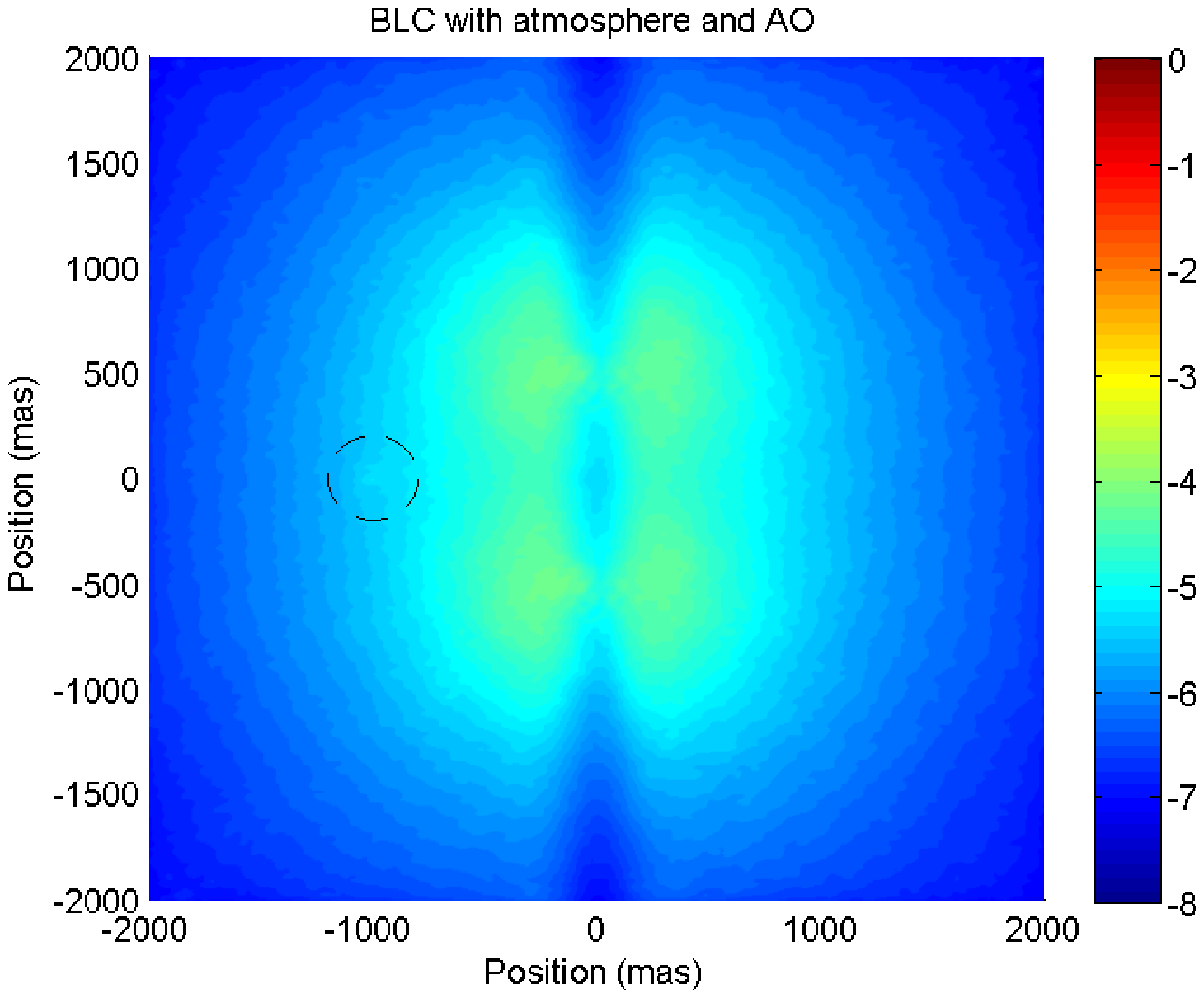}}
\caption{Performance of the four coronagraphs in the presence of atmospheric aberrations, wavefront control, and static error.  The intensity at each point is shown on a log scale. The companion in each image is at (-1000mas, 0mas).  The vertical band in the center of panel (d) is obscuration from the mask and no companions can be detected along that axis.}
\end{center}
\end{figure}

We examined the performance of four coronagraphic configurations, beginning with the case of no coronagraph in the system; the performance is shown in \myfig{m1}. The HiCIAO system currently uses a Lyot coronagraph for its observations; we looked at the performance with a standard round hard-edged mask doubled, with each mask centered at $\pm 500$\,mas.  At $1.65\mu$m, the mask diameter is $4.8\lambda/D$.  In addition, the Lyot stop is undersized to $80\%$ and the secondary is enlarged to $140\%$ \mycitep{Tam06}, to match the existing instrumentation.  The performance for this configuration is shown in \myfig{m2}.

\begin{figure}
\begin{center}
\includegraphics[width=3.1in]{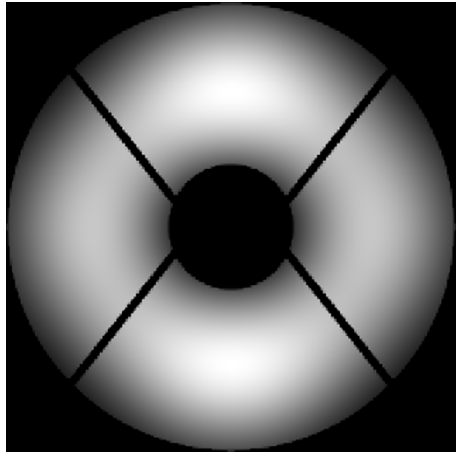}
\caption{The pupil of the apodized pupil Lyot coronagraph.}
\label{m3a}
\end{center}
\end{figure}

Using the techniques described in \mycitep{Sou09}, we designed an apodized pupil Lyot coronagraph for the Subaru pupil which explicitly incorporated the spiders; this pupil is shown in \myfig{m3a}.  Note that the inclusion of the spiders leads to breaking of circular symmetry in the apodization; in this case, there is higher throughput in the center of the upper and lower regions than in the side regions.  Again, the hard-edged image plane mask was doubled, and each mask centered at $\pm 500$\,mas in the image plane; the mask is identical to the one used for the Lyot coronagraph.  The Lyot stop is identical to the pupil in \myfig{subaruPupil}, as the eigenfunction equation underlying the APLC requires the Lyot stop match the pupil.  The performance of the coronagraph is shown in \myfig{m3b}.

\begin{figure}
\begin{center}
\subfigure{
\includegraphics[width=3.1in]{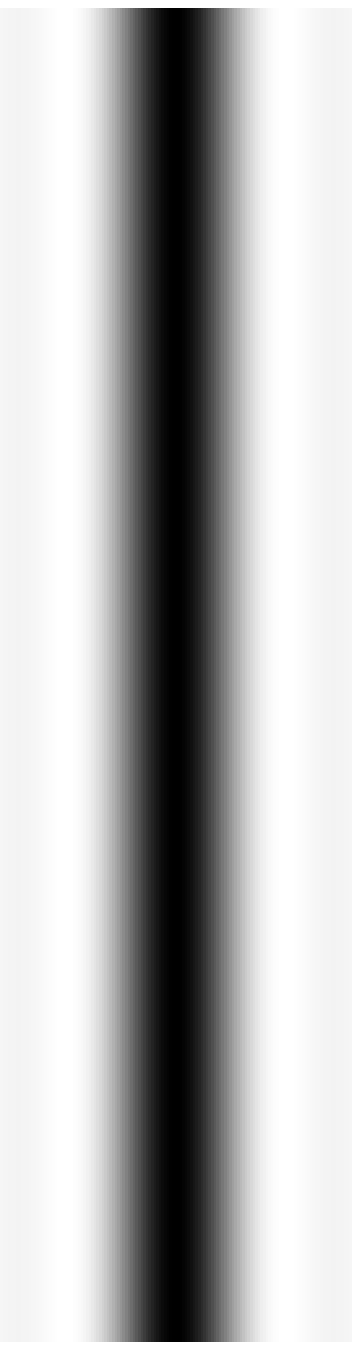}
\includegraphics[width=3.1in]{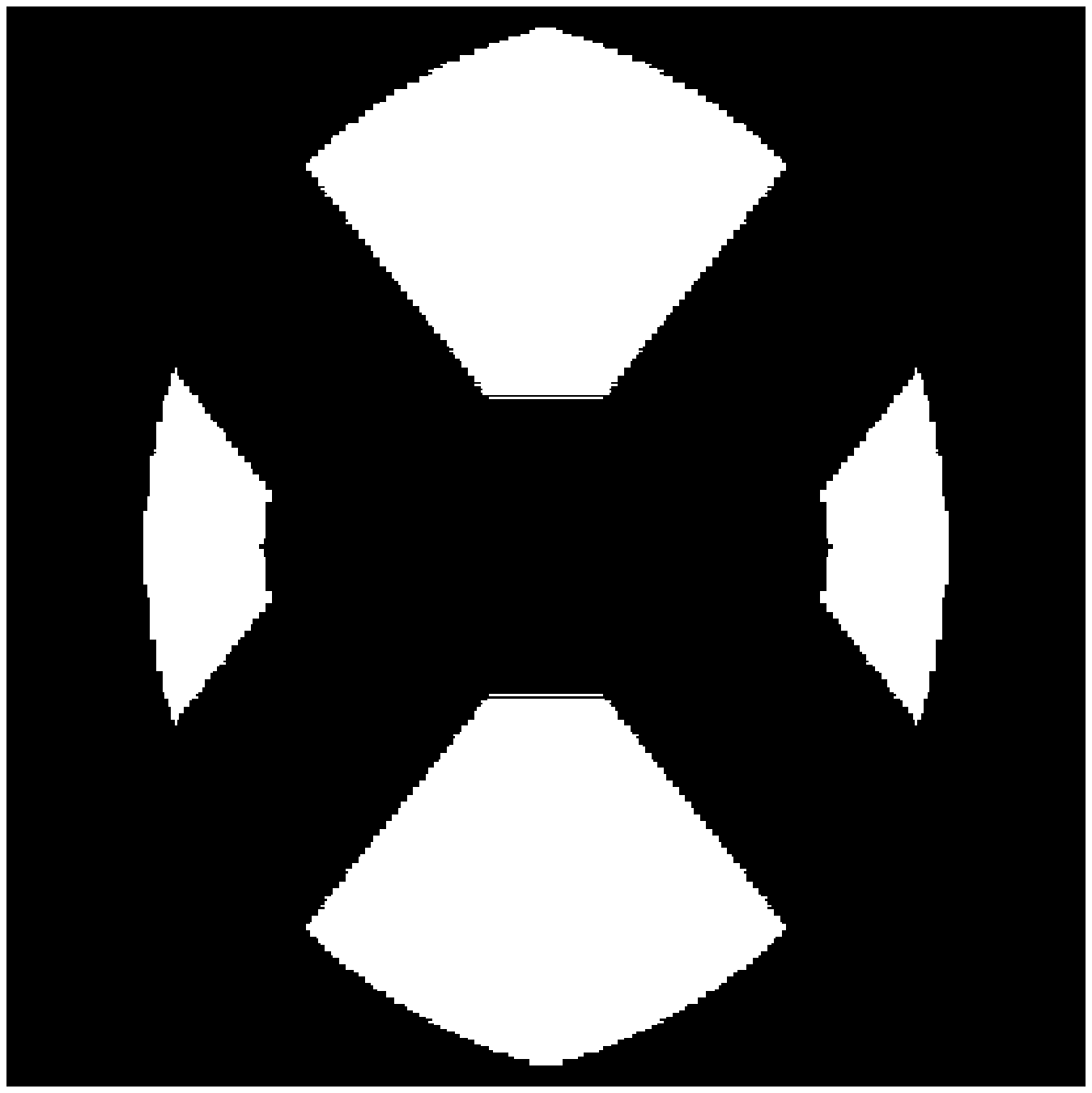}
}
\caption{\emph{Left.} The mask for the band-limited coronagraph.  \emph{Right.}  The Lyot stop for the band-limited coronagraph.}
\label{m4a}
\end{center}
\end{figure}

For the band-limited coronagraph, we aligned a 4th order linear mask along the direction of the two stars.  While higher-order designs exist \mycitep{Kuc05}, \mycitet{Cre07} shows that the 4th order mask generally outperforms them except in the presence of very high Strehl.  We use a modified Lyot stop, following \mycitep{Siv05}, to compensate for the large spiders present in the Subaru pupil.  The band-limited mask and Lyot stop used in our simulations are shown in \myfig{m4a}.  The Lyot stop is required to be quite severe to prevent the spiders from leaking into the image plane.  The performance of the mask is shown in \myfig{m4b}.

We find that the BLC and the APLC are both effective at suppressing the central cores of the PSFs and the spiders, although the throughput of the planet is reduced in both cases.  This reduced throughput is due to modifications to the system to eliminate spiders: the APLC has a throughput of $39.0\%$, with the reduction driven primarily by the apodizer, and the BLC has throughput of $35.36\%$, with the reduction driven primarily by the Lyot stop.  This result is not unexpected; \mycitet{Cre07} showed that in lower-Strehl cases, hard-edged masks will perform similarly to apodized image-plane masks, and the reduction in the opening of the Lyot stop between \myfig{subaruPupil} and \myfig{m4a} suggests a corresponding reduction in throughput.  As the apodizer is designed to concentrate the light into a central core, and the Lyot stop is not, the planet PSF is sharper for the APLC.

We note that rotation of the field will cause the spikes from the spiders to rotate, potentially obscuring the planet multiple times during the observation.  However, the BLC and the APLC prove effective at suppressing the diffraction from the spiders given their design considerations; this will be a major advantage to using either method on an obstructed aperture.

\section{Manufacturing considerations for the APLC}

We can conceptualize the masks as being two circles deposited on two discs of glass, each large enough such that the mask can be moved to any part of the first image plane without the edge of the glass appearing in the final image.  These discs would then be placed in a pair of X-Y actuated mounts so that the metal sides face each other, are separated by a very small amount, and bracket the image plane on either side.  For observations of single stars, the masks can be moved to coincide and perform as a standard APLC, or one moved aside completely if there is sufficient room.

Other physical arrangements should be possible, as well.  If the use of transmissive optics is excluded due to unwanted physical by-products such as ghosting or chromaticity, the masks can instead be etched to be freestanding with small spiders to hold them in place.  These spiders could then be included in the eigenfunction calculations, or simply included in the simulations as error sources.

The placement of the masks near the image plane will need to be precise to mm-scale, so one of the primary challenges in building this system is likely to be optomechanical---ensuring both masks remain in planes parallel to the image plane without hitting each other or becoming defocused.  The finite thickness of the masks is not likely to be a consideration, however, as existing freestanding image-plane masks can be sub-mm in thickness (e.g., 400$\mu$m for the ``bowtie'' masks used on the high-contrast testbed at Princeton \mycitep{Bel07}) as can coatings on coronagraph masks placed on glass (e.g., 200$\mu$m for 8th-order band-limited masks \mycitep{Lay05}).

In practice, it is better \emph{not} to optimize the APLC apodization for a specific tilt angle by creating an eigenfunction with two masks in the image plane.  An optimized APLC apodized mask will marginally improve the coronagraphic performance  for a binary system with a specific angular separation, but it would significantly degrade performance for stars that do not match to the apodization.  This configuration would require the apodizer to be replaced for every system.  Therefore, we believe it is better to compute the APLC apodization for an on-axis star and let Fourier optics solve the off-axis situation.  There are a number of methods by which the pupil apodization can be created.  For example, HEBS glass \mycitep{Boc08} and microdots \mycitep{Tho08, Mar09} are options being considered for APLCs on SPHERE and GPI, respectively, and shaped pupil APLCs offer another alternative for apodization \mycitep{Cad09b}.

One well-known shortcoming of the APLC is its chromaticity.  This could be compensated using standard linear optimization techniques to create an apodization that concentrates light and minimizes sidelobes across a designated bandpass.  Alternatively, the mask could simply be sized for the longest wavelength in the band, while shorter wavelengths create narrower PSFs whose cores are still blocked by the mask.  This may still scatter light in undesirable ways in the image plane.  A third option is the use of eigenvalues less than $\Lambda_0$, which may suppress more light off the central wavelength while still maintaining high suppression.  This method is employed in the APLC design for GPI \mycitep{Mac08}.  Some investigation would be required to determine which method is preferable, depending on the telescope arrangement.

\section{Tracking}

The majority of large telescopes today have obstructed apertures and alt-az mounts.  Field rotation will cause the binary pair to rotate in the image plane, with each star maintaining a copy of the telescope PSF.  There are two approaches that can be taken to compensate for the field rotation: simultaneous counter-rotation of the pupil and Lyot stop with the image rotator activated, and rotation of the image-plane masks with the image rotator deactivated.  Pupil rotation has been tested previously on sky: the pupil-tracking mode of the NACO instrument on the VLT rotates the entire instrument to keep the pupil fixed \mycitep{Kas09a, Tut10}.  While either is possible, rotating the image-plane masks may be preferred, as it minimizes the number of actuators required; the image-plane masks must be actuated regardless, to match the separation of the target binary.

The simplest way to place the masks in the image plane is with open-loop control--identify the image plane locations for the masks without the masks in, then move them into position and make science observations.  A fast steering mirror may be used to provide coarse correction, though as there are two sources being blocked independently, at least one mask will still have to be adjusted.   If this can be aligned with sufficient precision, the alignment procedure ends here.

If necessary, we suggest an additional closed-loop control to maintain mask alignment.  Closed-loop control will also be necessary if a system is run with the image rotator off, to ensure the masks rotate with the image plane.  One possible method of performing closed-loop control is to use flux in the image plane as a feedback signal.  Decentering the masks will tend to increase flux sharply, as this is equivalent to introducing tilt errors, which increases the flux leaking through an APLC \mycitep{Siv08}.  We note that pointing control using science camera imagery has already been demonstrated on-sky with a vector vortex mask, in the imaging of HR8799 from Palomar \mycitep{Ser10}, although the technique used to close the loop was not the same.

\begin{figure}
\begin{center}
\includegraphics[width=4in]{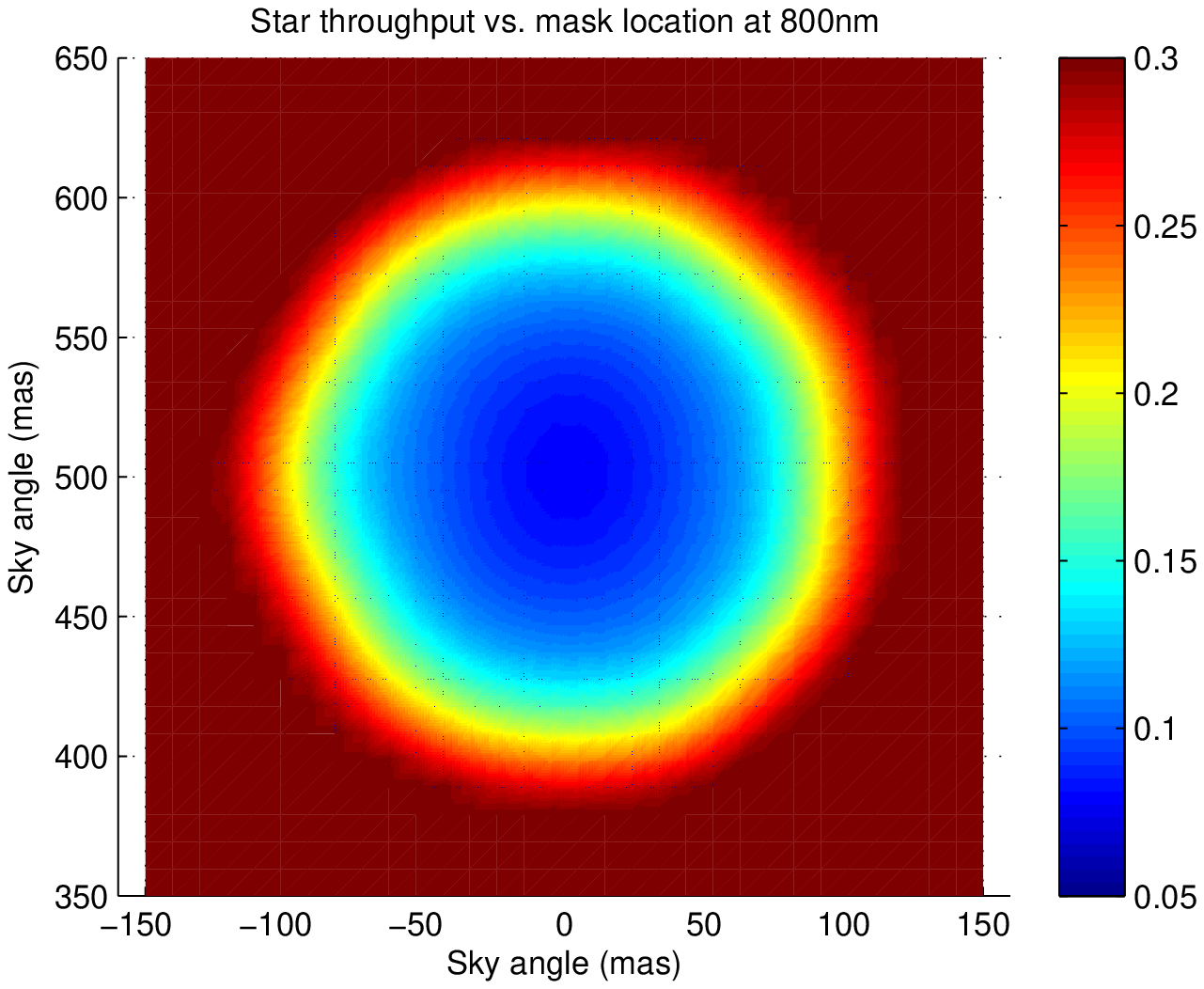}
\end{center}
\caption{Throughput at the second pupil plane, for various misalignments of one of the image plane masks.  The x- and y-coordinates show the location of the center of the mask; the ideal position is at $x_0 = 0$mas and $y_0 = 500$mas.  The throughput is summed over 2.5 seconds on wavefront and AO corrections.  The scaling is normalized such that the total energy from both stars combined is $1$ at the first pupil plane.}
\label{f2}
\end{figure}

The total energy in the pupil plane is shown in \myfig{f2} for a number of sky angles in the vicinity of a mask.  As shown, the region of suppression surrounding the correct alignment is distinct, though smeared by atmospheric errors.  Improved AO correction will improve the sensitivity.   Moving downward into this region would be relatively straightforward, if one mask is held fixed.   This has the added advantage that no additional optics would be required, as the control uses feedback from the the imaging camera.  Furthermore, we can iterate the positioning of the two masks if necessary, so only a single variable is used at a time.

We can then outline an observing procedure as follows: center the telescope between the two target stars, and obtain an estimate of the angular separation between them, and their angle relative to horizontal on the image plane.  This would be done with the masks displaced from the vicinity of the stars; the astrometric measurement creates an initial placement for the two masks in the image plane.  Closed-loop control, as outlined above, may proceed from there.

\section{Conclusions}
\label{sec:conclusion}

A sizable fraction of the overall planet population is expected to reside in binary systems; yet most current high-contrast facilities are not suitably equipped to observe such systems with coronagraphy.  We have presented a conceptual design for a APLC-based coronagraph which would allow faint companions to be seen around binary stars.  This design provides comparable throughput to observations using existing methods with band-limited coronagraphs, suppresses the diffraction spikes introduced by central obstructions and spiders, and allows observations without blocking edge-on systems.  We also outline a control scheme and observing procedure which would allow the coronagraph to lock onto the stars.  In particular, this would be a prime use for multi-object adaptive optics systems.

As a next step, we will examine mask tolerancing in simulation; a full closed-loop control system should be designed and closed-loop performance under noisy conditions should be simulated.  We would also hope to verify the performance experimentally with a small testbed model, and show the masks can track effectively.

\section*{Acknowledgments}

The authors would like to thank Sebastian Egner for graciously providing simulated wavefront data, Laurent Pueyo for useful discussions, and R{\'e}mi Soummer for providing code for creating apodizers for APLCs.  M.W.M. acknowledges support from NSF Astronomy \& Astrophysics Postdoctoral Fellowship under award AST-0901967.

\bibliography{refs}
\bibliographystyle{apj}

\end{document}